# Stable freestanding thin films of copolymer melts far from the glass transition

T.Gaillard[1], W.Drenckhan[1], C. Poulard[1], T.Voisin[1,†], C. Honorez[1], P.Davidson[1], M.Roché[1,‡]

1. Laboratoire de Physique des Solides, CNRS UMR 8502 & Université Paris Sud, Bâtiment 510, 91405 Orsay Cedex, France.
† Now at: Institut de Chimie de la Matière Condensée de Bordeaux, CNRS UPR 9048 & Université de Bordeaux, 33600 Pessac, France.
‡ Now at: Laboratoire Matière et Systèmes Complexes, CNRS UMR 7057 & Université Paris Diderot, 10 rue Alice Domon et Léonie Duquet, 75013 Paris, France.

## Abstract

Thin polymer films have attracted attention because of both their broad range of applications and of the fundamental questions they raise regarding the dynamic response of confined polymers. These films are unstable if the temperature is above their glass transition temperature $T_g$. Here, we describe freestanding thin films of centimetric dimensions made of a comb copolymer melt far from its glass transition that are stable for more than a day. These long lifetimes allowed us to characterize the drainage dynamics and the thickness profile of the films. Stratified regions appear as the film drains. We have evidence that the stability, thinning dynamics and thickness profile of the films result from structural forces in the melt. Understanding the key mechanisms behind our observations may lead to new developments in polymeric thin films, foams and emulsions without the use of stabilizing agents.

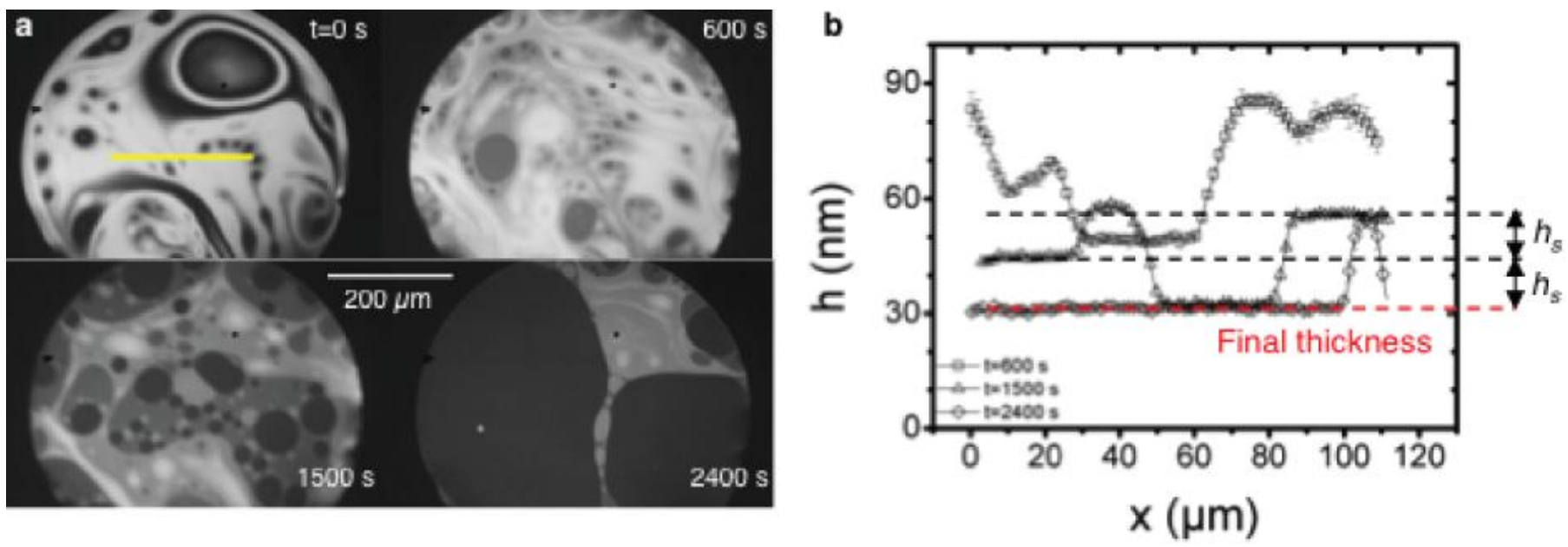

Many industrial and natural processes rely on the dynamic response of liquids confined to sub-micrometer dimensions. Examples include coating devices, foams and emulsions or lung airways[1]. Confinement can occur in "supported" conditions on a solid surface or in "freestanding" conditions, in which the liquid is confined by two free interfaces. Confinement effects become important when the distance between two interfaces separating the liquid from its environment is commensurate with a characteristic length scale of the liquid; the dynamics of the liquid may then change dramatically[2]. For example, thin films of complex fluids with thickness $h$ < 100 nm can have stratified thickness profiles whose stratum height is either correlated to the size of the molecules, as in the case of smectic films[3–5] or to the characteristic dimensions of aggregates/meshes formed by molecules in solution[6–10]. Another manifestation of confinement effects is the dependence of the glass transition temperature $T_g$ of polymeric materials on film thickness $h$ [11].

The response of polymeric materials to confinement has been investigated mostly in the supported case for thicknesses $h$ < 100 nm (see ref. 12 for a review). Besides a better understanding of the relationship between $T_g$ and $h$, questions such as the stability of thin polymer films against dewetting [13–15] or the structuration of block copolymers were tackled [16–18]. In contrast, studies of freestanding thin polymer films are scarce [12,19–25], more so in the case of copolymers [26,27]. Like their supported counterparts, the stability of these films decreases as temperature increases well above $T_g$.

Here we describe observations made for freestanding thin films of submicrometer thickness drawn from silicone-based, comb-like copolymer melts far from their glass transition. In contrast to the hitherto studied freestanding thin polymer films, the films we study are characterized by lifetimes on the order of one day while reaching thicknesses on the 10-nm scale. Besides their phenomenal stability, these films display a stratified thickness profile. We discuss the mechanism that may explain the resistance of these films to rupture as well as their stratification. We show that the latter originates most likely from structural forces in the liquid that could also explain the stability of the films.

We used a commercially available comb-copolymer (DBP-732, Gelest Inc., molecular weight $M_w \approx 20\ 000$ g.mol$^{-1}$, polydispersity index = 5) that is molten at room temperature and at least 60ºC above its glass transition temperature (Fig. S1 in Supplementary Materials). The copolymer consists of a poly(dimethylsiloxane) (PDMS) backbone with approximately $N_b = 110$ siloxane groups of which 27 carry statistical poly(ethylene glycol)/poly(propylene glycol ) (PEG/PPG) copolymers (Fig. 1a). The side-chains are $N_s = 10$ units long ($x = 4$ PEG units and $y = 6$ PPG units on average), $CH_3$-terminated, and connected to the backbone by a $(CH_2)_3$ group. We measured the rheological properties of the melt using an Anton Paar MCR 301 rheometer with the cone-plate geometry. Rotational tests showed that the melt viscosity is constant, $\eta = 2.7$ Pa.s, over the range of shear rates $0.1 < \dot{\gamma} < 1000$ s$^{-1}$ (Fig. S2 in Supplementary Materials).

Vertical thin films of DBP-732 are formed at room temperature using a square 30-mm wide frame made of 160-µm-thick nylon fiber frame held inside a metal structure (Fig. 1b), in analogy to Mysels' historic soap film experiments[28]. The film is entirely pulled out of a vessel containing the melt at constant velocity $U = 5$ mm.s$^{-1}$ and then left to drain after removal of the vessel. Figure 1c shows an image of a DBP-732 film after thirty hours. Colors result from the interference pattern of white light reflected at normal incidence by the film. Reflected light is analyzed to extract thickness maps by collecting it sequentially with three narrow band-pass filters ($\lambda_1 = 450$ nm, $\lambda_2 = 550$ nm, $\lambda_3 = 660$ nm) and processing the acquired signals using a MatLab code based on Scheludko's micro-interferometric method [29,30].

DBP-732 films remain stable over timescales on the order of $10^5$ s**,** allowing us to study their drainage dynamics. Figure 2a shows a typical image sequence taken at $\lambda$ = 660 nm over a duration of 1600 min after the film has been pulled out. While the film is initially quite smooth (Fig. 2a, $t$ = 60 s), recirculation flows grow across the whole film and a flow along the vertical nylon fibers brings liquid from the bottom of the film to its top (Fig. 2a, $t = 2\times10^3$ s and $3\times10^4$ s). The origin of such flows are often indications of the presence of surface tension gradients [31,32], but we have no proof of such an assumption yet for this system. As drainage progresses, recirculating flows vanish, and the top of the film becomes smooth ($t = 10^5$ s). Common black films grow at the top of the film long before bursting: the film thickness $h$ has finally decreased to a few tens nm. In this region, the film is stratified (Fig. S3 in Supplementary Materials) much like other complex-fluid thin films[6–10]. We found the height of strata to be multiples of $\Delta h_v = 16 \pm 1$ nm, as discussed in more detail later**.** We sometimes observed stratification in regions whose thickness is on the order of 500 nm**.**

Figure 2b shows that the vertical thickness profiles of DBP-732 evolves with time like the profiles of vertical soap films draining under gravity[28,33,34]. We investigated the ability of existing soap-film drainage models to capture the thinning dynamics $h(z,t)$ we observed. First, the long lifetime of the films led us to test Reynolds' drainage model with rigid interfaces[28,35] that predicts the evolution

$$h(z,t) = \frac{h_0}{\sqrt{1 + \frac{4}{3} a(z)(t - t_0)}} \qquad (1)$$

with $= \rho g z h_0^2 / \eta L^2$ , $\rho$ the density of the liquid, $g$ the gravitational acceleration, $z$ the position on the film, $h_0$ the initial thickness at time $t_0$, $\eta$ the viscosity of the liquid, and $L$ the height of the film. Figure 2c shows that a fit of Eq.1 to the experimental data with the initial thickness equal to the thickness of the film right after the removal of the vessel, $h_0$ = 447 nm, and $\alpha$ as an adjustable parameter works very well. However, the value we obtained for $\alpha$ suggest that $\eta = 10^{-4}$-$10^{-3}$ Pa.s, a value that is at least three orders of magnitude smaller than the bulk viscosity of the liquid we measured. The opposite situation of drainage between free interfaces is also unable to capture the trends that we report, as the thinning dynamics we observe is not exponential[33]. This last result also highlights the fact that viscous resistance linked to a high bulk viscosity does not explain the long lifetime of the films, in contrast with thin polymer films made of very viscous silicone oils ($\eta = 10^3$ Pa.s in the latter case compared to $\eta$ = 2.7 Pa.s here)[36]**.** Thus thin-film drainage models are unable to rationalize the dynamics of our films. Finally, as film thicknesses are larger than 100 nm during most of the drainage process, we expect DLVO forces between the interfaces to be negligible. The observation of stratification suggests that significant structural forces not accounted for in the drainage model we used exist in DBP-732 that may affect thin-film dynamics.

We investigated the existence of structural forces using horizontal films that we studied using a thin-film pressure balance[37–39] (TFPB) designed to work with viscous fluids. Thickness measurements are performed as in the case of vertical films. At $t$ = 0 the pressure difference between the liquid and the surrounding nitrogen is increased to $10^3$ Pa. Figure 3a shows images of the film during its thinning, indicating that the drainage process may be divided into two stages. During the first stage, the film thickness is inhomogeneous and varies smoothly across the film (Fig. 3a, $t$ = 0 s). When the film reaches thicknesses on the order of 70 nm, strata appear at the boundaries between well-defined zones of constant thickness (Fig. 3a, t = 600, 1500 and 2400 s).

As thinning progresses, strata appear all over the film and the liquid drains out of the film layer by layer (Fig. 3b). The minimum thickness of the film before breakup is $h_{min}$ ~ 30 nm, and we never observed black films with $h < h_{min}$ even close to bursting. This is in contrast with the vertical film, where $h_{min}$ ~ 16 nm. A statistical treatment of the thickness maps of both, vertical and horizontal films, shows peaks in the distribution of thicknesses (Fig. 3c). These peak values were reproducible over many film occurrences and a linear fit to the data shows that the average height difference between two strata in both types of films is $\Delta h \approx 16 \pm 1$ nm (Fig. 3d). Similar observations have been reported for thin films of almost glassy diblock copolymers when film thickness becomes commensurate with the molecular size[26,40].

We investigated the bulk molecular features of the copolymer using small-angle X-Ray scattering (SAXS) performed on the SWING beam line at SOLEIL ($\lambda$ = 1.08755 Å, sample-detector distance 1 m). We observed an isotropic scattering motif for DBP-732 (Fig. 4a and inset) with a peak at a scattering vector $q_c = 0.040 \pm 0.007$ Å$^{-1}$ that indicates a bulk characteristic length $l_c \approx 15.7 \pm 2.7$ nm similar to the stratum height $\Delta h = 16 \pm 1$ nm that we reported earlier. Pattern isotropy is consistent with the absence of a measurable signal when performing birefringence analysis on the sample.

Comparison with a theoretical estimate of the size of a comb polymer gives indication of the nature of $l_c$. The macromolecular conformation of branched polymers depends on the polymerization degrees $N_b$ and $N_s$ of the backbone and side-chain, respectively, and on the number of backbone units $n$ between consecutive side-chain grafting points. We are in the limit of short non-Gaussian side-chains ($N_s < 100$) [41,42] and of a Gaussian backbone [43]. The spacers $l_s$ between branching points are approximately four siloxane units long on average, i.e. $l_s \approx 1.3$ nm considering that the distance between two silicon atoms in the backbone is given by $2l_{Si\text{-}O}$ with $l_{Si\text{-}O} = 0.164$ nm[44] and assuming that the monomers are linearly arranged. In comparison, the non-Gaussian side-chains have an approximate full-extension length $l_{sc}$ ~ 4.5 nm[45]. As a consequence, the side-chains see each other and they minimize their interactions by forcing the backbone to stretch. Thus we hypothesize that the copolymers form dense combs [41,42]. Moreover, since we consider melt properties, we assume that individual chains are in a θ-solvent. Assuming that only steric interactions matter in setting the macromolecular conformation[42] and neglecting bond length differences, the radius of gyration $R_g$ of the copolymer may be estimated[42]

$$R_g \approx \left( \frac{N_{tot}}{N_s} \frac{l_s}{D} \right)^{3/5} D, \qquad (2)$$

with $N_{tot} = N_b + N_s$ the total number of monomers in the comb and $D$ the diameter of the comb. In our case, $N_{tot} \approx 460$, $N_b = 13$ (including the alkyl linker), and D = $2l_{sc} \approx 9$ nm. Calculation gives $R_g \approx 24$ nm, a value of the same order of magnitude than $l_c$. Thus $l_c$ is likely the radius of gyration of DBP-732. We note that the geometrical features of DBP-732 molecules indicate that they must have a non-negligible persistence length and should therefore be prolate rods rather than spheres.

We therefore propose the following hypothesis (Fig. 4b) to explain our observations: while the bulk polymer does not show any ordering transition (as can be seen in Fig. S1 in the Supplementary Materials), the presence of the interfaces and the induced confinement lead to a stratified organization the characteristic length of which is the radius of gyration of the individual molecules. As the surface tension of DBP-732 $\gamma = 20.7 \pm 0.2$ mN.m$^{-1}$ (Fig. S4 in Supplementary materials) is close to that of pure PDMS of similar chain length[46], $\gamma_{PDMS} = 20.1 \pm 0.2$ mN.m$^{-1}$, the first layer at the interface is probably composed of molecules which orient their PDMS backbone to the surrounding air[47]. Such a molecular conformation may help imposing a preferential order of the rod-like copolymers close to the interface thus encouraging stratification. This mechanism may also explain the observation of stratification in films several hundred nm thick. Once the film thickness is of the order of a few times the radius of gyration, the isotropic bulk liquid gives way to a layered system. How exactly the polymers organize in these layers remains an open question and will be subject to future investigations.

The occurrence of stratification in freestanding copolymer films is known to be related to an increase of film stability[27]. In their study, Croll and Dalnoki-Veress suggest that the layered structures formed by some copolymers when confined in freestanding films increase the resistance of these films to thermal fluctuations and DLVO interactions when compared to other homo- and copolymers that do not organize. DBP-732 might show excellent stability for similar reasons, especially since the thickness of the film reaches a minimum value $h_{min} \sim 16$ nm that would limit the importance of short-range and thermal fluctuations.

In conclusion, we described the ability of a comb-like copolymer melt far from its glass transition temperature to form day-long stable thin films. We discuss the drainage of these films and identify structural forces in the copolymer melt as the origin of both stratification and the long lifetime of the films. Many questions remain open such as the role of chain flexibility, molecular structure (backbone and side-chain lengths, chemical composition of the side-chains,…) and the exact organization of the polymers in the film. Preliminary results indicate that similar copolymers with 10 to 20 units in the backbone are unable to form stable films. Thus molecular structure is a crucial parameter and its control would allow tuning the stability of thin copolymer films far from their glass transition. In turn, foams and emulsions could be obtained without a stabilizing agent (Fig. S5 in Supplementary Materials), opening new perspectives in fields such as cosmetics or material design.

## Acknowledgements

We thank J. Venzmer, M. Zeghal, C. Gay, E. Rio, L. Champougny, B. Scheid, D. Langevin, L. Léger, H. Plantefève, F. Restagno, P. Guénoun, R. von Klitzing, P.-A. Albouy and A. Cagna for fruitful discussions.

## Funding

This project was funded by an ERC Starting grant (agreement 307280-POMCAPS). We received additional financial support from TECLIS and BPI France.

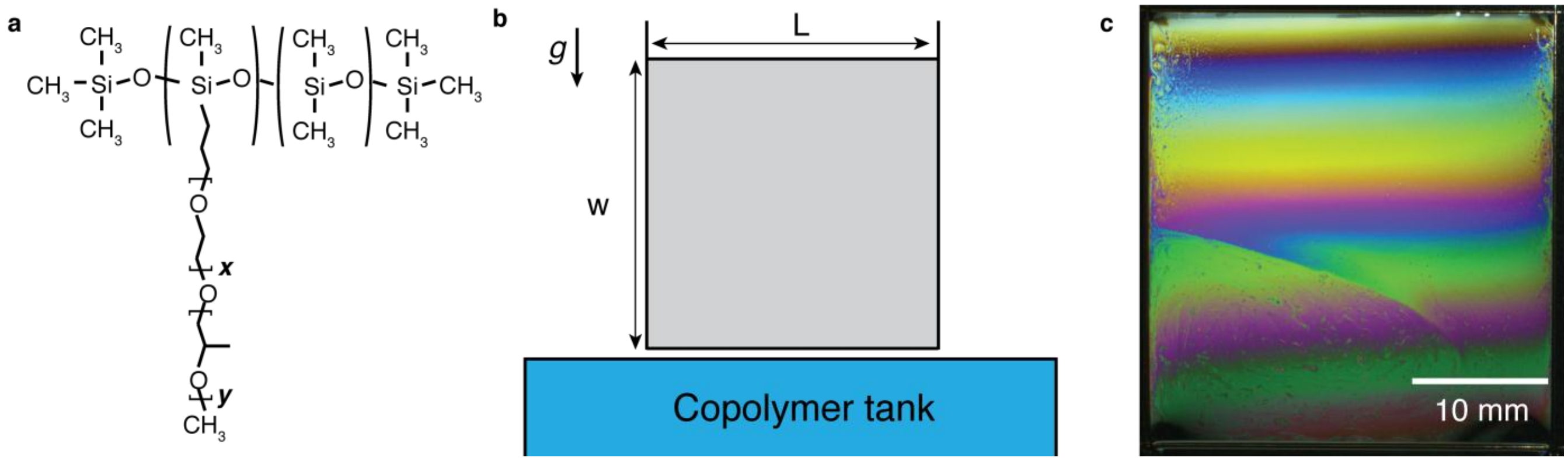

Figure 1: a) Molecular structure of DBP-732. x and y denotes the numbers of PEG and PPG units in the side-chains respectively. b) Sketch of the pulling film setup. L= w = 30 mm. Films are formed by pulling the frame out of the copolymer tank at a speed velocity U = 5 $mm.s^{-1}$. c) Image of a vertical thin liquid film of DBP-732, t = 1800 min.

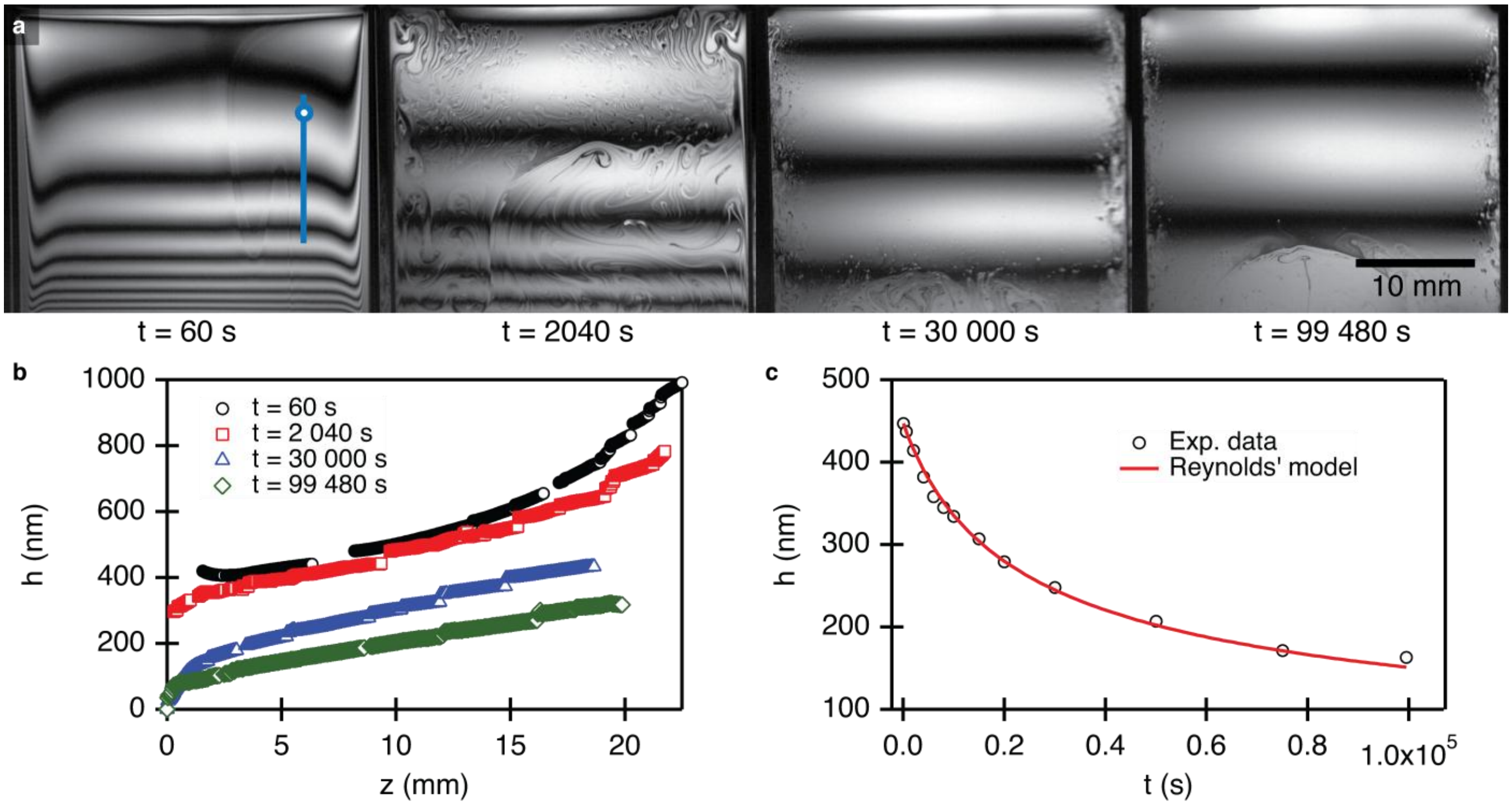

Figure 2: a) Monochromatic image sequence of a vertical DBP-732 film taken at λ = 660 nm. b) Time evolution of film along blue line on image a at t = 60 s. z=0 is taken has the top of the film. c) Film thickness h versus time t at z = 6 mm under the top nylon fiber (Mean over the disk on image a, t = 60 s). Circles: experimental results. Red line: fit with Reynolds' model for drainage between two immobile surfaces.

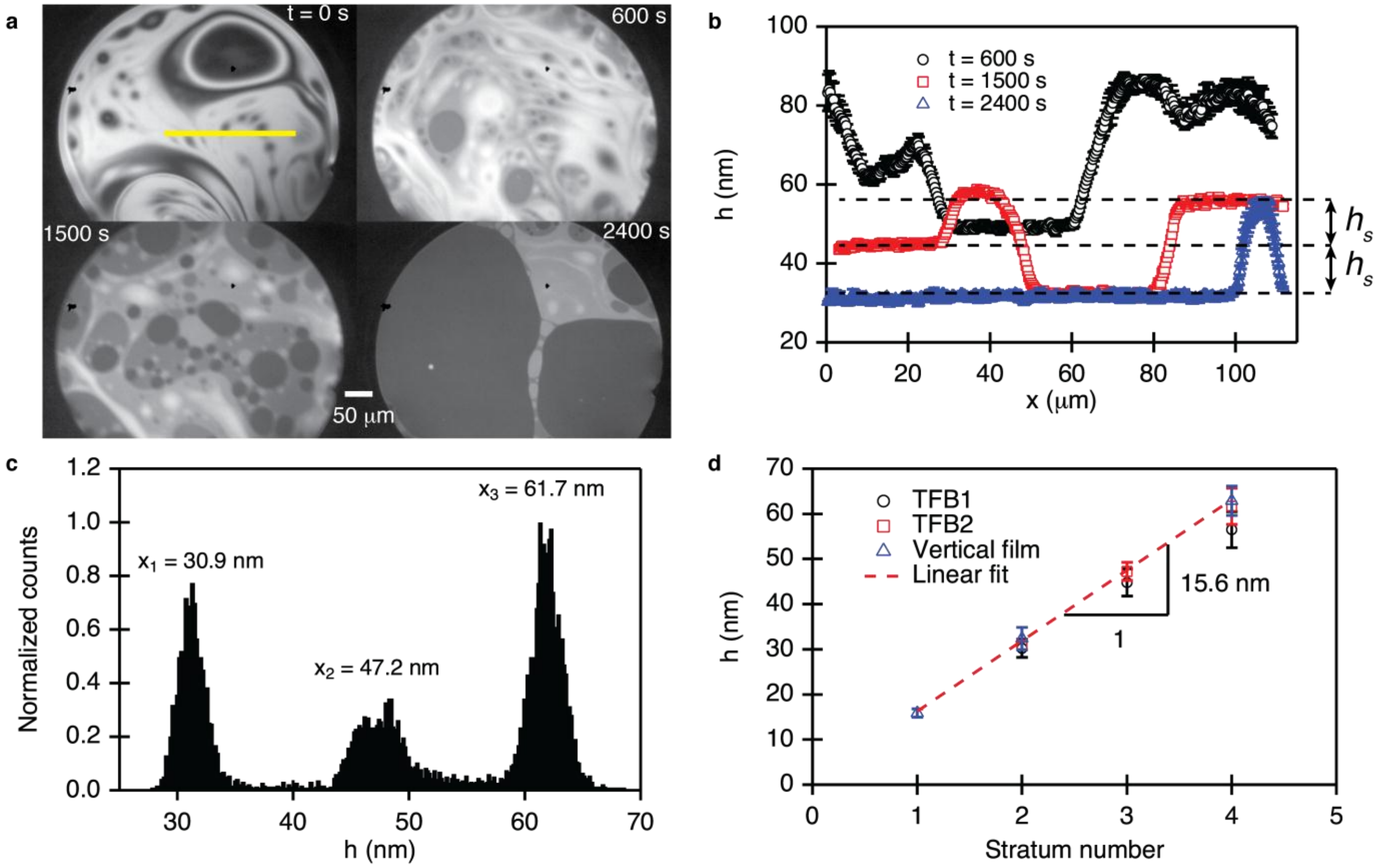

Figure 3: a) Image sequence of a horizontal draining film. b) Film thickness profiles along the yellow line on image a, t = 0 s. c) Thickness distribution across horizontal films. The $x_i$ values indicate the location of the peaks. Data acquired on five films under identical experimental conditions. d) Thicknesses of the homogeneous areas that appear simultaneously on horizontal films (open black circles and open red squares), vertical films (open blue triangles). Dashed line : linear fit to the data.

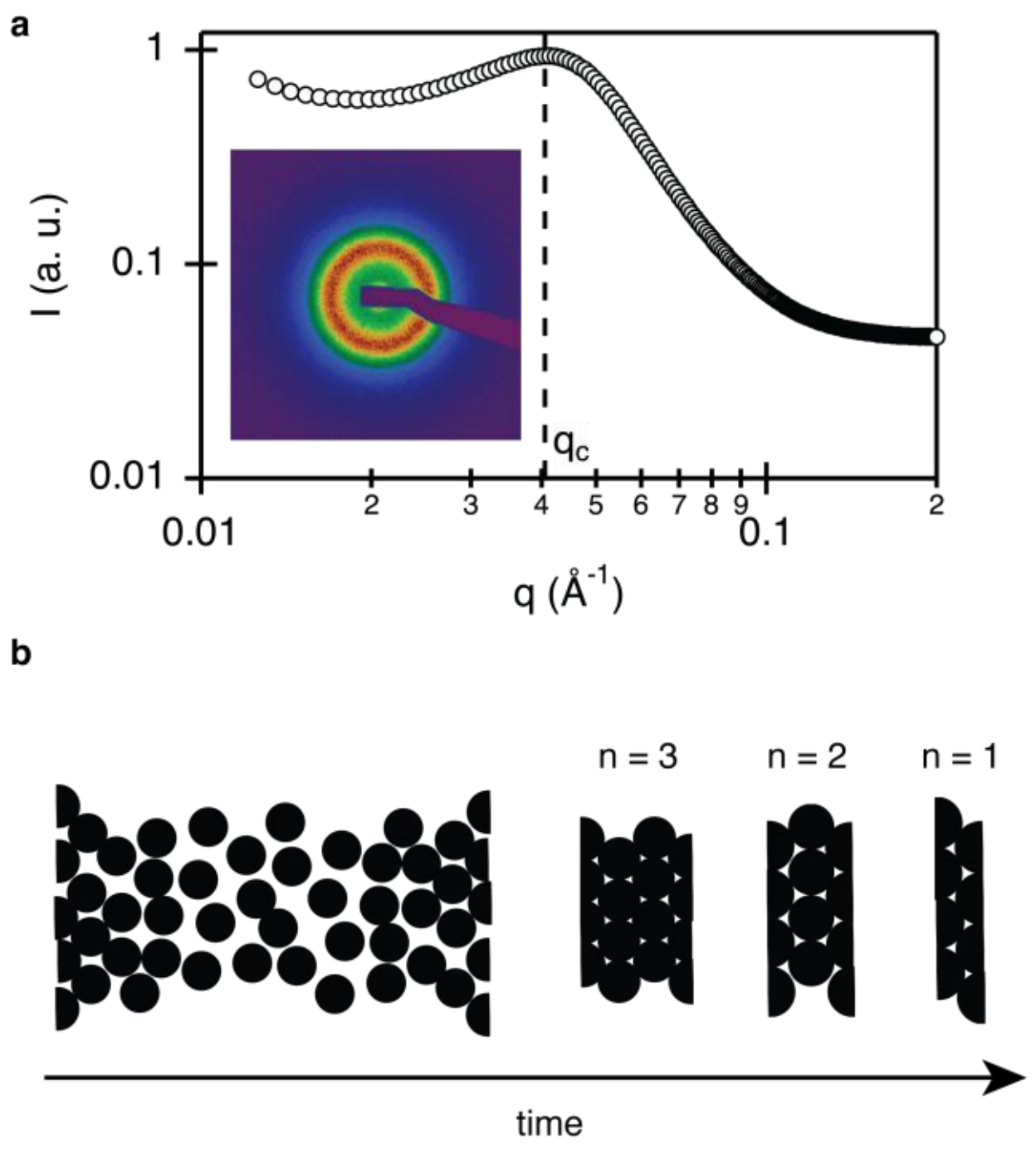

Figure 4: (a) SAXS scattering intensity in the bulk of DBP-732. The black experimental dots are angular averages of all the radial intensity profiles seen on the scattering image (inset). (b) Suggested structure of the film as drainage proceeds. n denotes the number of molecular diameters.